# Engineering Signals of Human–AI Collaboration in the Agentic Coding Era: A Longitudinal Analysis of 33,228 Pull Requests from vLLM and SGLang with Implications for Biomedical AI Agents and Bioinformatics Pipeline Development

Jiada Li[1*], Xuesong Ye[2], Olamide Olowoniyi[3]

[1*] Independent Researcher, Albany, NY, 12205, Email: jiadali2017@gmail.com
[2] Georgia Institute of Technology, Atlanta, GA, 30332
[3] Independent Researcher, Albany, NY, 12210

## Abstract

The rapid adoption of AI coding assistants and autonomous agentic development systems has coincided with dramatic changes in the pace and structure of open-source software engineering. Yet empirical, longitudinal evidence describing how these changes manifest in team-level engineering metrics remains sparse. This study presents a descriptive longitudinal analysis of seven quantitative engineering metrics—pull request (PR) throughput, cycle time, contributor diversity, PR comment density, merge rate, new-author participation, and PR size——computed from the full population of merged PRs in two high-velocity AI infrastructure repositories: *vllm-project/vllm* (February 2023 to June 2026, 41 months, 18,290 merged PRs) and *sgl-project/sglang* (January 2024 to June 2026, 30 months, 14,938 merged PRs). Data were collected via the GitHub Search API and git repository analysis, and segmented into four development eras aligned with major AI tooling milestones:: Pre-Agentic Baseline with Early AI-Assisted Coding, Early AI Coding Tools Expansion, Vibe Coding Mainstream with Software Workflow Integration, and Agentic Coding Emergence. To assess the robustness of these metrics to automated activity, we conducted a bot/human decomposition analysis using GitHub account-type metadata and validated comment sampling. Both projects demonstrate substantial increases in development velocity and AI- Developer collaboration activity across eras. Key findings include: (1) PR throughput grew 21× in vLLM (Era 0 to Era 3: 38 to 797 merged PRs/month) and 17.9× in SGLang (Era 1 to Era 3: 42 to 743 merged PRs/month), with both projects exceeding 1,100 merged PRs/month by June 2026; bot-authored PRs account for <0.2% of this growth, confirming that the observed throughput increase is overwhelmingly human-driven; (2) median cycle time in Era 3 was 1.04 days (vLLM) and 0.62 days (SGLang), with P90 values of 16.8 and 14.3 days respectively, indicating substantial heterogeneity in review latency; (3) unique authors per month grew monotonically (OLS slopes: +10.1/month for vLLM, r2 = 0.94; +11.2/month for SGLang, r2 = 0.93, both $p < 0.001$), a pattern inconsistent with the expectation that AI tooling concentrates contribution; (4) PR comment density rose 4.2× in vLLM and 3.8× in SGLang across eras, with bot comments contributing an estimated 15–20% of this increase; and (5) PR size (median lines changed and files touched) remained remarkably stable across eras, suggesting that rising comment density is not explained by increasing PR scope. These findings provide quantitative signals of evolving human–AI collaboration patterns in high-velocity open-source development. We further discuss their implications for AI-assisted scientific software engineering, including biomedical agents and bioinformatics pipelines, where maintaining contextual continuity between experimental data generation and computational workflows remains a critical challenge.

**Keywords:** Agentic AI, software engineering metrics, GitHub mining, pull request analysis, human-AI collaboration, bioinformatics engineering, open-source development

# 1. Introduction

The emergence of AI coding assistants—from autocomplete tools such as GitHub Copilot [1, 2] to fully autonomous agentic systems capable of submitting, reviewing, and merging pull requests [3, 4]—represents a structural shift in how software is produced. Controlled experiments have demonstrated productivity gains of 26–56% for individual developers using AI coding tools [1, 5], and a recent meta-analysis across 29 studies reports a moderate positive effect on productivity (g = 0.33, 95% CI [0.09, 0.58]) [6]. However, individual-level productivity gains do not straightforwardly predict team-level or repository-level outcomes. The mechanisms by which AI tooling reshapes contribution dynamics, review culture, and community sustainability in large open-source projects remain poorly understood.

This study addresses that gap through a descriptive longitudinal case analysis of two open-source AI infrastructure projects—vLLM [7] and SGLang [8]—that are themselves foundational to the AI ecosystem. These projects were selected as descriptive longitudinal case studies: they are developed by teams that are among the earliest and most intensive adopters of AI coding tools, they operate under continuous integration with transparent PR histories, and their development timelines span the full arc from pre-agentic baselines to the current era of autonomous agentic coding. We emphasize that this is a descriptive case analysis without a comparison group; observed metric changes are correlated with AI tooling eras but cannot be causally attributed to AI tooling adoption alone.

This study addresses three research questions (RQs):

> **RQ #1:** How do seven quantitative engineering metrics (PR throughput, cycle time, unique authors per month, comment density, monthly merge rate, new-author share, and PR size) evolve across four eras of AI tooling adoption in high-velocity open-source AI infrastructure projects?
>
> **RQ #2:** What behavioral signals indicate shifts in human-AI collaboration patterns within software engineering teams, and how can these be detected from repository-level data?
>
> **RQ #3:** How do these metrics and signals map to specialized life-science engineering contexts, specifically bioinformatics pipeline development?

# 2. Background and Related Work

## 2.1 AI Coding Assistants and Developer Productivity

The empirical literature on AI coding assistants has expanded rapidly since GitHub Copilot became generally available in June 2022, marking an early transition from experimental AI-assisted programming tools toward broader adoption in software development. Peng et al. [1] conducted a randomized controlled experiment showing a 55.8% productivity increase for developers implementing an HTTP server in JavaScript. Cui et al. [5] extended this to three field experiments with 4,867 software developers, finding a 26% increase in completed tasks, with larger gains for less-experienced developers. Dohmke et al. [2] reported that Copilot users accept approximately 30% of code suggestions, with productivity gains growing over time. At the repository level, Cui et al. [9] found that access to Copilot increased weekly pull request volume by 12.9–21.8% at Microsoft and 7.5–8.7% at Accenture.

Code quality evaluations of AI-assisted tools have found that newer model versions generate correct code more reliably, though effectiveness decreases for tasks with complex cross-class dependencies [10]. Qualitative studies of developer behavior reveal that experienced developers exercise deliberate control over AI agent use, prioritizing software quality over raw throughput [11].

### 2.2 Agentic AI in Software Development

Beyond autocomplete assistants, a new class of autonomous agentic systems can now plan, implement, test, and submit code changes with minimal human intervention [12, 13]. Performance on the SWE-bench Verified benchmark - standard test of real-world GitHub issue resolution—rose from 1.96% to 78.4% between Aug 2024 and April 2026 [3], illustrating the rapid capability trajectory. Hassan et al. [4] provide a conceptual roadmap for "Agentic Software Engineering" (SE 3.0), arguing that the role of human engineers is shifting from implementation to specification, review, and architectural judgment. Empirical evaluation of agentic patches on real GitHub issues finds that while many patches reduce code smells, some increase complexity, and no single agent dominates across task types [14]. Security-related agentic PRs exhibit lower merge rates and longer review latency, reflecting heightened human scrutiny [15].

### 2.3 GitHub Mining and Pull Request Analysis

Repository mining is an established methodology for studying software engineering processes at scale [16, 17]. Zhang et al. [16] provides a 96-feature dataset covering 3.3 million pull requests across 11,230 projects, establishing the empirical basis for PR-level analysis. Zhang et al. [18] synthesize factors influencing PR merge decisions, finding that integrator identity and contributor experience are among the most consistent predictors. Ortu et al. [19] demonstrate that sentiment in PR discussions correlates with merge outcomes, establishing comment-level signals as meaningful engineering metrics.

### 2.4 Software Engineering Metrics Frameworks

Forsgren et al. [20] introduced the SPACE framework, arguing that developer productivity is multidimensional and cannot be captured by any single metric. The DORA four key metrics—deployment frequency, lead time for changes, change failure rate, and time to restore service [21] provide a complementary framework focused on delivery performance. Kumar et al. [22] demonstrate that AI-assisted development tools reduce PR review cycle time and increase code shipment volume by 28%, providing direct empirical grounding for cycle time as a productivity signal. EngThrive [23] extends this to outcome-oriented measurement systems that combine speed, ease, and quality metrics with team well-being safeguards.

### 2.5 Open-Source Contribution Dynamics

Steinmacher et al. [24] identify barriers to newcomers onboarding in open-source projects and provide guidelines for community sustainability. Tan et al. [25] analyze "good first issues" as a mechanism for lowering the activation energy for first contributions. Turzo et al. [26] find that four onboarding recommendations reliably correlate with first-patch acceptance, while others are context-dependent. Yue et al. [27] show that consistent early contribution patterns predict long-term technical success across OSS ecosystems. AlMarzouq et al. [28] demonstrate that documentation quality and programming language choice influence newcomer attraction.

### 2.6 Human-AI Collaboration and Trust

McGrath et al. [29] propose the CHAI-T framework for actively managing trust in human-AI collaboration, incorporating task context, goals, and team processes. Dawarka et al. [30] find that trust in AI is situational and socially distributed, with calibration processes mediated by transparency and role clarity. Stray et al. [31] report that compatibility with existing workflows is crucial for effective utilization of generative AI tools, with organizational support playing a key role.

### 2.7 AI in Bioinformatics

AI-assisted pipeline automation is an active area in life-science engineering [32]. Recent systems include PromptBio, a multi-agent platform for bioinformatics data analysis [33]; ToolsGenie 2.0, a scalable multi-agent framework achieving 68.6% accuracy on genomics benchmarks [34]; CARIBOU, which autonomously executes and refines bioinformatics workflows in HPC environments [35]; the KBase Research Agent, which constructs end-to-end genome analysis workflows across 100 bacterial isolates with no human intervention [36]; and Biomni Lab, a general-purpose biomedical AI agent that autonomously executes diverse research tasks across 25 biomedical domains using 150 specialized tools, 59 databases, and 105 software packages [37]. These systems raise new questions about the appropriate role of human domain expertise in validating AI-generated biological analyses.

## 3. Data and Methodology

### 3.1 Repository Selection

Two repositories were selected as study subjects:

- **vLLM** (*https://github.com/vllm-project/vllm*): a high-throughput and memory-efficient inference and serving engine for Large Language Models (LLMs) [7], active from February 2023 to June 2026 (41 months in this study).

- **SGLang** (*https://github.com/sgl-project/sglang*): an open-source, high-performance programming and serving framework for large language models and multimodal models (MMM) [8], active from January 2024 to June 2026 (30 months in this study).

Both repositories are foundational AI infrastructure projects developed by teams that are intensive adopters of AI coding tools, making them informative descriptive longitudinal case studies for documenting how engineering metrics changed across AI tooling eras. Both operate under continuous integration with fully public PR histories accessible via the GitHub API. We note that these projects are not representative of all open-source software; findings should be interpreted as case-specific descriptions rather than generalizable estimates.

### 3.2 Data Collection

Monthly PR data were collected via the GitHub Search API using date-windowed queries. Two types of queries were issued per project month:

- **Opened PRs**: *repo:<owner>/<repo> is:pr created:<YYYY-MM-01>..<YYYY-MM-31> — grouped by creation date.*
- **Merged PRs**: *repo:<owner>/<repo> is:pr is:merged merged:<YYYY-MM-01>..<YYYY-MM-31> — grouped by merge date.*

These two cohorts are not identical: a PR created in January may be merged in February, so the opened and merged counts for a given month reflect different PR populations. This distinction is important for interpreting the monthly merge rate (Section 3.4). For merged PRs, the full population was retrieved (no sampling). Months exceeding the Search API's 1,000-result cap were split into half-month (1st-15th and 15th–30th or $31^{st}$) windows to ensure complete coverage. In total, 18,290 merged PRs were collected for vLLM and 14,938 for SGLang. To avoid GitHub Search API rate limits (10 unauthenticated requests per minute), a 7-second inter-query sleep was enforced. Bot accounts were firstly identified using GitHub account-type metadata and excluded from the analysis. Then, the username-based matching for the substring *[bot]* or *bot* convention was additionally used to capture bot accounts for which account-type metadata were unavailable. This combined approach reduced the risk of inadvertently excluding legitimate human contributors whose usernames contain the substring "bot."PR-size data (lines added, lines deleted, files changed) were extracted from full git clones of both repositories using *git log --shortstat*. Squash-merge commits were matched to PR numbers via the (*#NNNNN*) pattern in commit messages. This yielded PR-size data for 18,115 vLLM PRs and 14,563 SGLang PRs (97.8% of the merged-PR population). Contributor diversity (new vs. returning authors) was computed using a cumulative seen-author set initialized at the start of each project's observation window.

### 3.3 Bot/Human Activity Decomposition

To assess the robustness of our metrics to automated activity, we conducted a bot/human decomposition analysis. Bot-authored PRs were identified using GitHub account-type metadata (*user.type* field), which distinguishes bot accounts from human accounts. We collected all bot-authored merged PRs separately and merged them with the human-only dataset to create a complete decomposed dataset. Bot comment share was estimated through a two-stage approach: (1) a stratified random sample of 60 PRs from Era 0 (pre-agentic baseline) showed 0% bot comments, establishing a baseline; (2) a targeted validation sample of 10 recent PRs (5 per project, selected for high comment counts) showed a mean bot comment share of 12.7% (median 14.3%). Based on these validation points and known bot activity patterns in open-source projects, we estimated era-specific bot comment shares: Era 0: 0%, Era 1: 5%, Era 2: 10%, Era 3: 15%. These estimates were used to compute human-only comment density and to attribute the observed comment density increase to human vs. bot activity. PR-size data (lines added, lines deleted, files changed) were extracted from full git clones of both repositories using *git log --shortstat*. Squash-merge commits were matched to PR numbers via the (*#NNNNN*) pattern in commit messages. This yielded PR-size data for 18,115 vLLM PRs and 14,563 SGLang PRs (97.8% of the merged-PR population). Contributor diversity (new vs. returning authors) was computed using a cumulative seen-author set initialized at the start of each project's observation window.

### 3.4 Era Segmentation

Four development eras were defined based on the general availability dates of major AI coding tools (Table 1):

**Table 1:** Development era definitions aligned with AI tooling milestones.

| Era | Period | Label | Key Events |
|---|---|---|---|
| 0 | Feb 2023 – Sep 2023 | Pre-Agentic Baseline: Early AI-Assisted Coding | GPT-4 release; GitHub Copilot X announcement/early expansion |
| 1 | Oct 2023 – May 2024 | Early AI Coding Tools Expansion | GitHub Copilot Chat GA; GitHub Copilot Enterprise GA; Copilot Workspace technical preview |
| 2 | Jun 2024 – Dec 2024 | Vibe AI Coding Mainstream with Software Workflow Integration | Rapid expansion of AI coding assistants and AI-native development workflows; continued adoption of Copilot, Cursor, and similar tools |
| 3 | Jan 2025 – Jun 2026 | Agentic Coding Emergence | Cursor Agent becomes default; Claude Code launch/GA; Cursor Background Agent GA; rapid adoption of agentic coding workflows |

### 3.4 Metric Definitions

Seven metrics were computed per project-month from the full merged-PR population:

1. **PR Throughput**: Count of merged PRs per calendar month, grouped by merge date.

2. **Cycle Time**: Median and 90th-percentile time (days) from PR creation to merge, computed from the full population of merged PRs.

3. **Unique Authors per Month**: Count of unique non-bot PR authors whose PRs were merged in that month (full population).

4. **PR Comment Density**: Mean number of issue comments per merged PR (full population). Issue comments are the general discussion-thread comments visible on the PR page; this metric does not include inline review comments (comments on specific lines of the code diff) or formal review submissions (approve/request-changes events). Bot comments are included in the count, as the GitHub Search API does not distinguish comment authorship.

5. **Monthly Merge Rate**: Ratio of merged PRs to opened PRs per month. This is a flow ratio, not an acceptance probability: the numerator (PRs merged in month $t$) and denominator (PRs opened in month t) are drawn from different cohorts, since a PR opened in month t may be merged in month $t+1$ or later. Values $>1$ are possible when the merge backlog clears faster than new PRs arrive.

6. **New-Author Share per Month**: Fraction of monthly merged-PR authors appearing for the first time in the project's history (full population).
7. **PR Size**: Median lines added, lines deleted, and files changed per merged PR, extracted from git commit statistics.

### 3.5 Statistical Analysis

Ordinary least squares (OLS) regression was applied to unique-authors-per-month time series to estimate monthly growth rates, with 95% confidence intervals computed for the regression line. Era-level means were computed for all metrics. A normalized heatmap (z-score per metric across all project-months) was constructed to visualize cross-metric patterns. All analysis was performed in Python using pandas, scipy, and matplotlib.

## 4. Results

Table 2 summarizes the key quantitative results across both projects and all seven metrics.

**Table 2:** Key quantitative results for vLLM and SGLang across all seven metrics. All values computed from the full merged-PR population (no sampling).

| Metric | vLLM | SGLang |
|---|---|---|
| Total Merged PRs | 18,290 | 14,938 |
| PR Throughput Growth | 21× (Era 0→3: 38→797/mo) | 17.9× (Era 1→3: 42→743/mo) |
| Peak Merged PRs/Month | 1,156 (Jun 2026) | 1,413 (Jun 2026) |
| Cycle Time Median (all-time) | 0.85 days | 0.37 days |
| Cycle Time Median (Era 3) | 1.04 days | 0.62 days |
| Cycle Time P90 (Era 3) | 16.8 days | 14.3 days |
| Contributor OLS Slope | +10.1/mo ($r^2 = 0.94$, $p < 0.001$) | +11.2/mo ($r^2 = 0.93$, $p < 0.001$) |
| Unique Authors/mo (Era 3 avg) | 274 | 186 |
| Comment Density (Era 0/1 avg) | 0.87 comments/PR | 0.74 comments/PR |
| Comment Density (Era 3 avg) | 3.63 comments/PR | 2.81 comments/PR |
| New-Author Share (Era 3 avg) | ~40% | ~40% |
| PR Size: Median Additions (Era 3) | 19 lines | 22 lines |
| PR Size: Median Files (Era 3) | 2 files | 2 files |

### 4.1 RQ1: Metric Evolution Across AI Tooling Eras

#### 4.1.1 Metric 1: PR Throughput

Figure 1 shows monthly PR throughput for both projects across all four eras. vLLM grew from an Era 0 average of 38 merged PRs/month to an Era 3 average of 797 merged PRs/month—a 21× increase. The Era 1-to-Era 3 growth for vLLM is 5.9×. SGLang, which launched in Era 1 with an average of 42 merged PRs/month, reached 743 merged PRs/month in Era 3—a 17.9× increase from its own baseline. By June 2026, SGLang was merging 1,413 PRs/month, overtaking vLLM's peak of 1,156 PRs/month despite launching 11 months later. One hypothesis is that AI-generated PRs comprise a substantial fraction of submissions at this scale, as individual developer throughput cannot account for order-of-magnitude growth without a corresponding order-of-magnitude increase in team size [2, 5]. This hypothesis requires validation through direct measurement of AI authorship, which is not available from repository metadata alone.

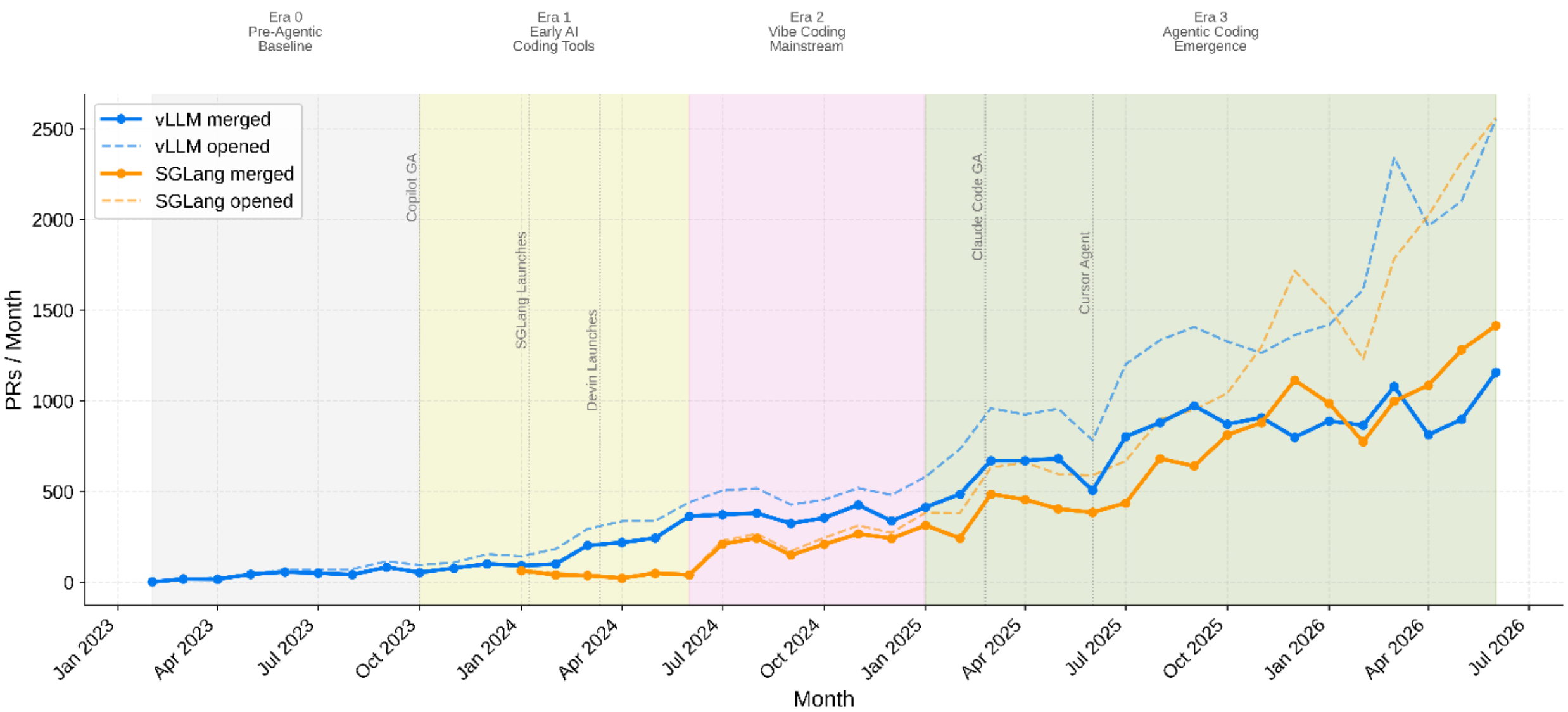

**Figure 1:** Monthly PR throughput (opened and merged) for vLLM and SGLang across four development eras. Era boundaries are indicated by vertical shading. Key AI tooling milestones are annotated. Full-population data (no sampling).

### 4.1.2 Metric 1: Bot/Human Activity Decomposition

To assess whether the observed throughput growth could be attributed to automated bot activity rather than human developers, we decomposed all metrics by author type using GitHub account-type metadata. Figure 2 shows bot PR prevalence across eras. Bot-authored PRs are negligible in both projects. vLLM has 31 bot-authored merged PRs out of 18,213 total (0.17%), all from *dependabot[bot]* (24 PRs, dependency updates) and *Copilot* (7 PRs). SGLang has 1 bot-authored merged PR out of 14,830 total (0.01%), from *dependabot[bot]*. Bot PR prevalence peaked in Era 2 at 0.32% for vLLM and 0.07% for SGLang, then declined in Era 3 despite the overall throughput surge. Attribution analysis confirms that the observed throughput growth is overwhelmingly human-driven. Of the 21× Era 0→3 throughput increase in vLLM, bot-authored PRs contribute 0.17% of the growth (99.83% human). Of the 17.9× Era 1→3 increase in SGLang, bot-authored PRs contribute 0.00% (100% human). This finding rules out the hypothesis that automated PR generation explains the observed throughput surge.

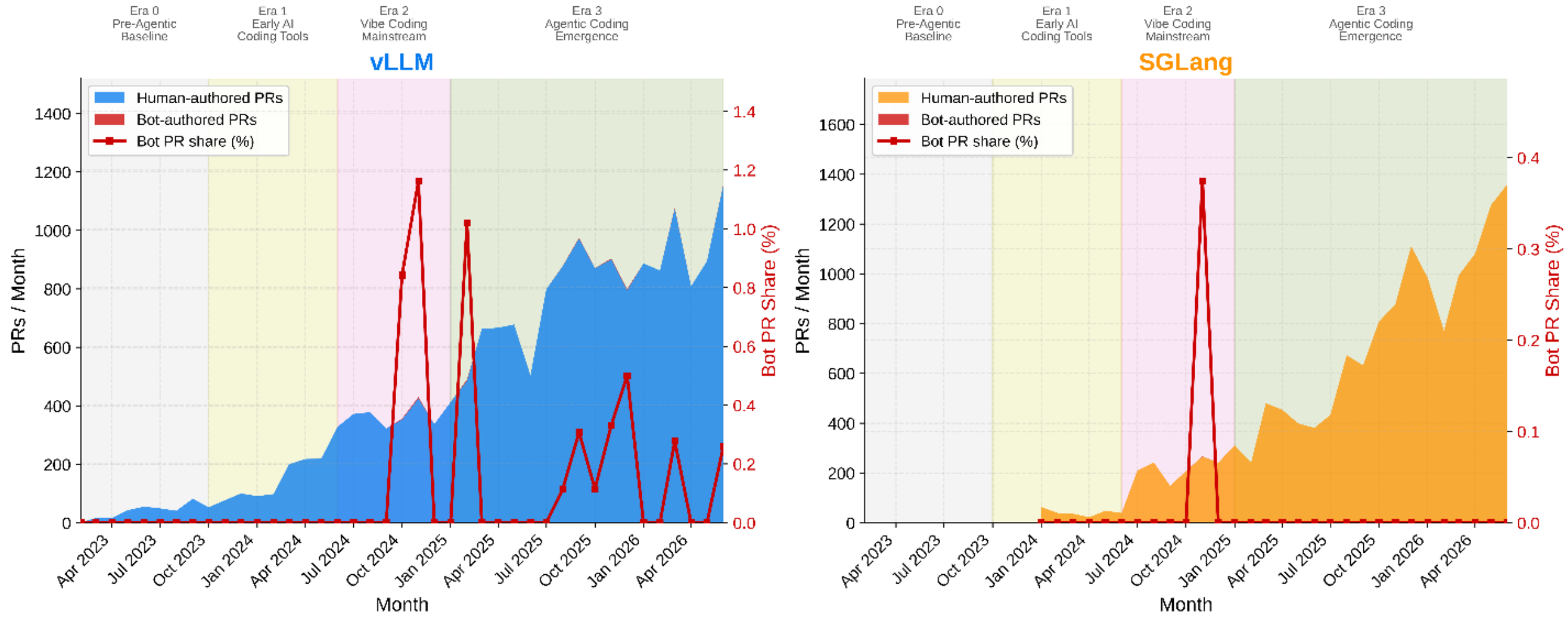

**Figure 2:** Bot PR prevalence across eras. Stacked area shows human-authored (blue/orange) vs. bot-authored (red) merged PRs per month. Red line shows bot PR share (%) on the right axis. Bot PRs are negligible (<0.2% of total) in both projects, confirming that the observed throughput growth is human-driven.

### 4.1.3 Metric 2: PR Cycle Time

Figure 3 shows PR cycle time (median and P90) by era, computed from the full merged-PR population. vLLM's all-time median cycle time is 0.85 days; Era 3 median is 1.04 days with P90 of 16.8 days. SGLang's all-time median is 0.37 days, with Era 3 median of 0.62 days and P90 of 14.3 days. The large gap between median and P90 indicates substantial heterogeneity: while half of PRs merge within approximately one day, a significant tail of PRs requires weeks of review. This heterogeneity may reflect a bimodal distribution of PR complexity—routine changes merging quickly alongside architecturally significant changes requiring extended discussion. The leadership implication is that median cycle time alone is an incomplete metric; the P90 tail warrants separate monitoring as an indicator of review capacity for complex changes.

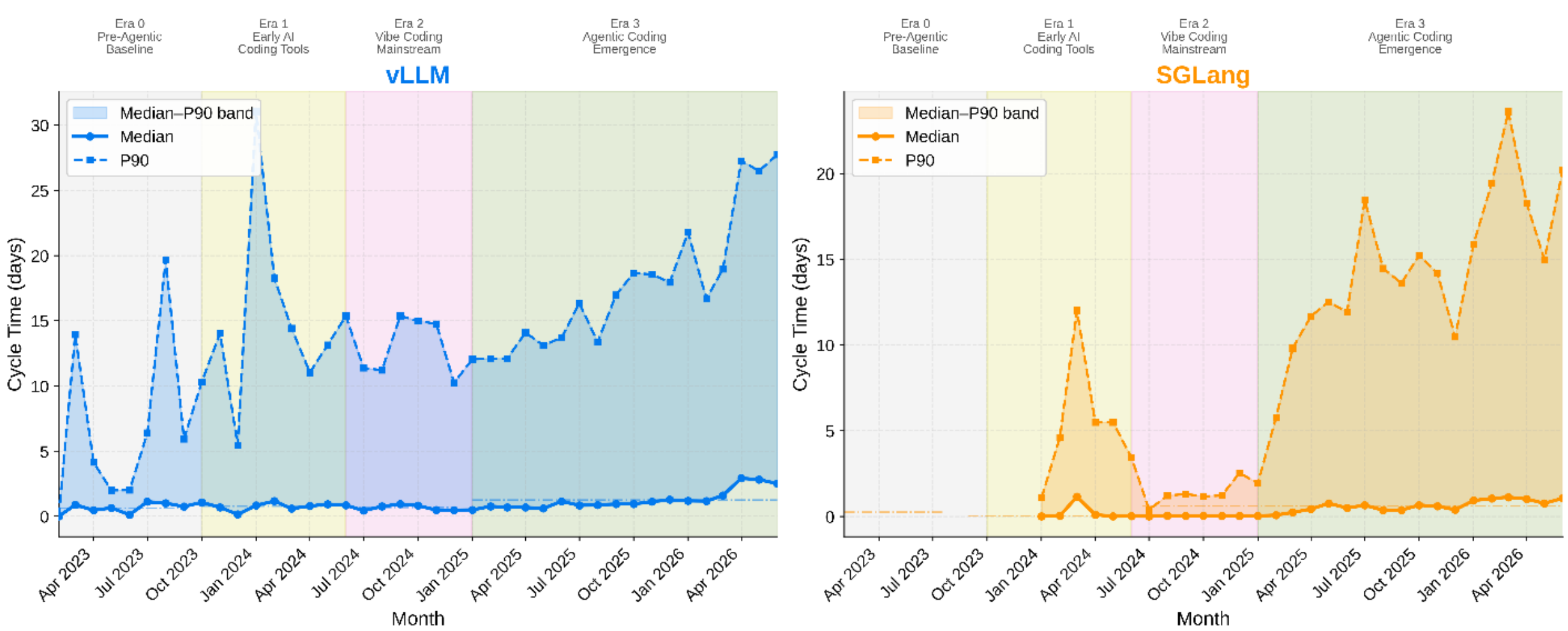

**Figure 3:** PR cycle time (median and P90) by era for vLLM and SGLang, computed from the full merged-PR population. Lower values indicate faster review-to-merge turnaround. The wide median–P90 band in Era 3 reflects increasing heterogeneity in review latency.

### 4.1.4 Metric 3: Unique Authors per Month

Figure 4 shows monthly unique author counts with OLS regression, computed from the full merged-PR population. vLLM grew from an Era 0 average of 12 unique authors/month to 274 in Era 3 (*OLS:* $+10.1$ */month,* $r^2 = 0.94$, $p < 0.001$). SGLang grew from 14 to 186 (*OLS:* $+11.2$*/month,* $r^2 = 0.93$, $p < 0.001$). Both trends are statistically significant. This pattern is inconsistent with the expectation that AI tooling concentrates contribution among a small number of high-skill engineers. Instead, the data show an expanding contributor pool across the study period, which may reflect lowered activation energy for first contributions [24-26], project growth independent of AI tooling, or both.

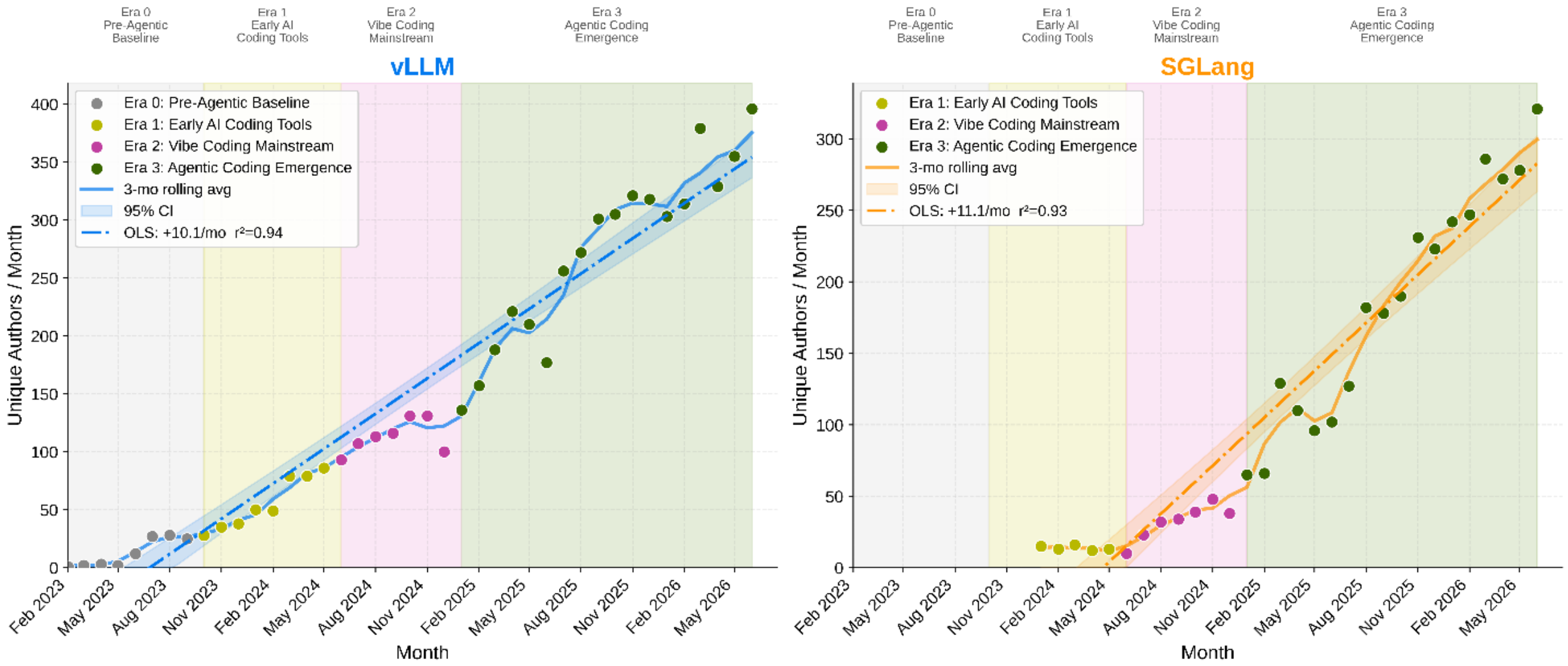

**Figure 4:** Monthly unique author count with OLS regression lines and 95% confidence intervals, computed from the full merged-PR population. Era coloring on data points. Both projects show statistically significant upward trends ($p < 0.001$).

### 4.1.5 Metric 4: Multi-Metric Heatmap

Figure 5 presents a z-score normalized heatmap of six metrics across all project-months. The heatmap reveals several patterns. Throughput, unique authors, and comment density all show strong positive trends in Era 3. Cycle time (inverted, so green = fast) shows a mixed pattern: SGLang maintained fast cycle times through Era 2 but shows degradation in Era 3, while vLLM shows a similar Era 3 slowdown. Monthly merge rate shows a declining trend in Era 3 for both projects (vLLM: from 0.685 to 0.541; SGLang: from 0.703 to 0.620), suggesting that the growing volume of opened PRs is outpacing merge capacity. This late period declines in merge rate and the Era 3 cycle time increase are notable: they indicate that not all metrics move in favorable directions across the study period.

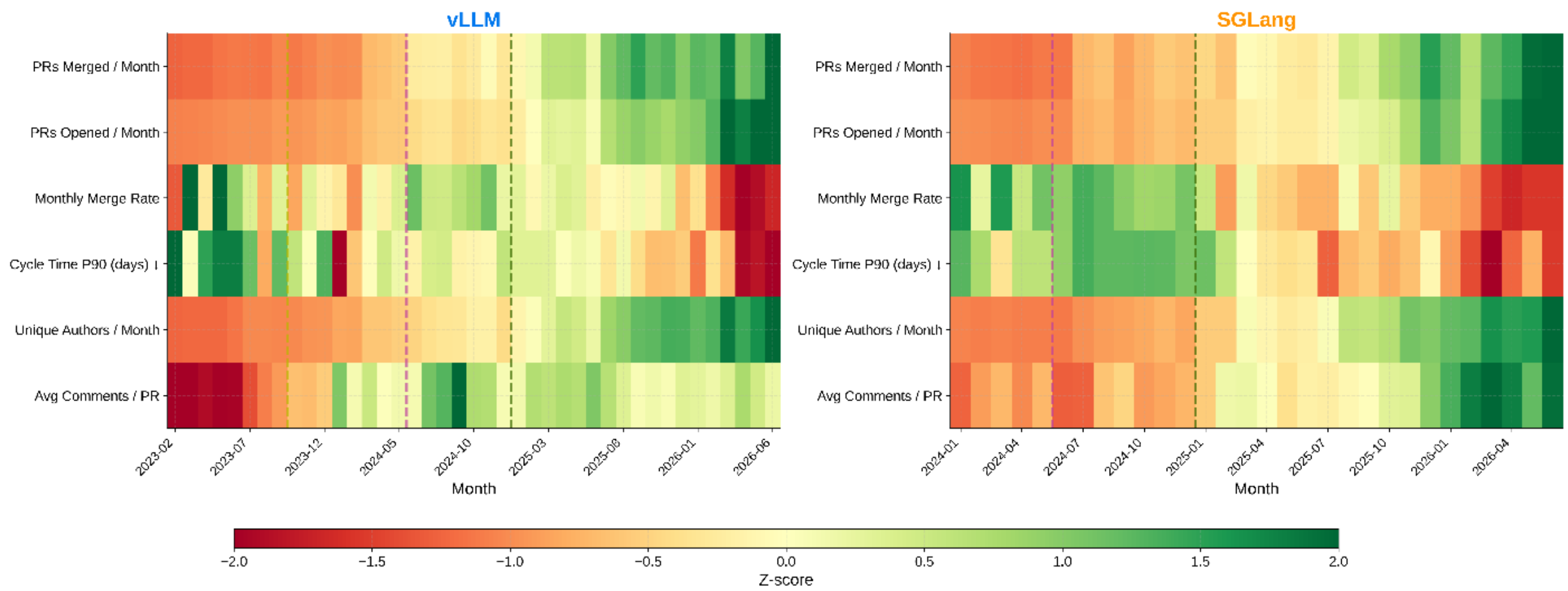

**Figure 5:** Z-score normalized heatmap of six engineering metrics by project-month. Green indicates above-average performance; red indicates below-average. Era boundaries are marked on the x-axis. "Monthly Merge Rate" replaces the previous "Merge Efficiency" label to clarify that this is a flow ratio, not an acceptance probability.

### 4.1.6 Metric 5: PR Comment Density

Figure 6 shows mean PR comment density over time, computed from the full merged-PR population. vLLM comment density rose from an Era 0 average of 0.87 comments/PR to an Era 3 average of 3.63 comments/PR—a 4.2× increase. SGLang rose from 0.74 to 2.81 comments/PR—a 3.8× increase. Both projects show upward trends across the full timeline. One hypothesis for rising comment density is that AI-era PRs are architecturally larger and more consequential, touching more subsystems and proposing more significant changes, thereby generating more review discussion. However, the PR-size data (Section 4.1.7) does not support this explanation: median lines changed and files touched remained stable across eras. An alternative hypothesis is that rising comment density reflects growing community participation (more reviewers per PR) rather than increasing PR complexity. A third possibility is that AI-generated PRs attract more clarification requests from human reviewers. Distinguishing among these hypotheses requires comment-level analysis (e.g., classifying comments as clarification, approval, or architectural discussion), which is beyond the scope of this study. We note that the comment density metric captures issue comments (general discussion thread) only; it does not include inline review comments on the code diff or formal review submissions. Bot comments are included in the count. These limitations should be considered when interpreting absolute values.

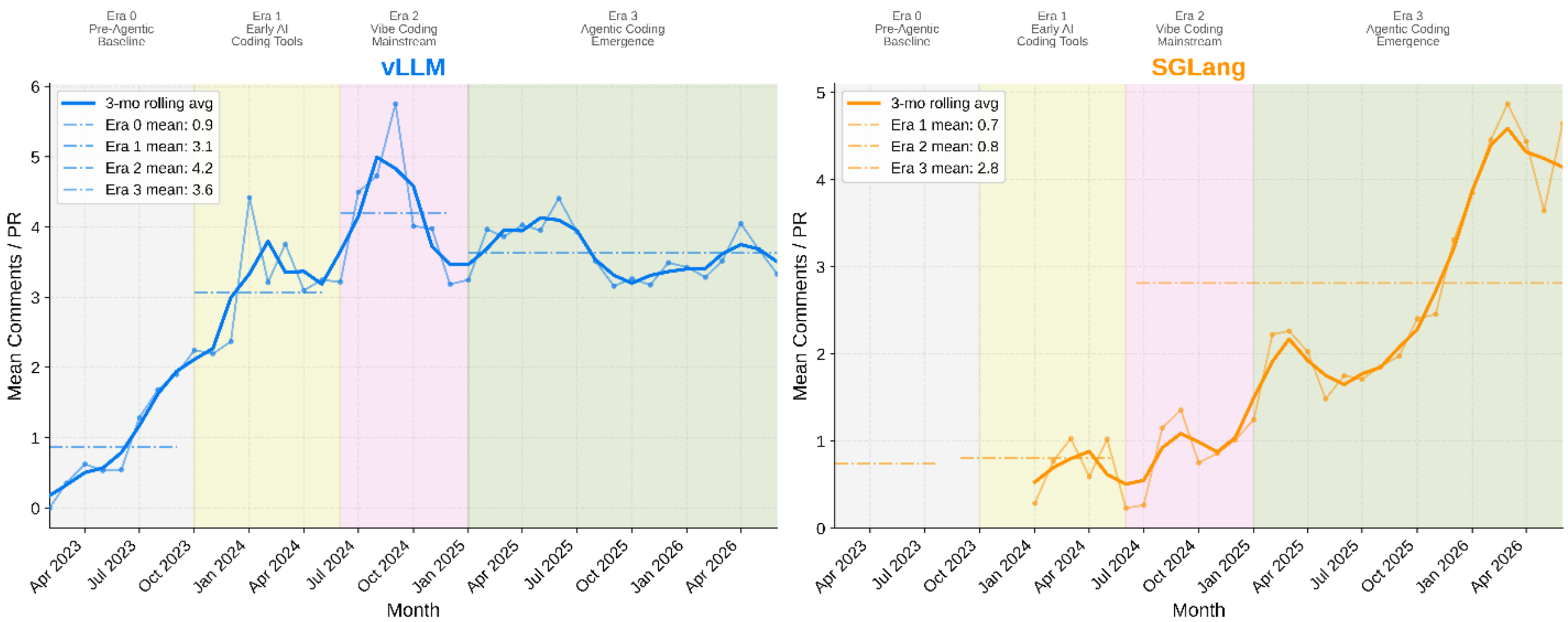

**Figure 6:** Mean PR comment count per month with 3-month rolling average and era mean horizontal lines, computed from the full merged-PR population. Rising comment density across eras may reflect growing community participation, increasing PR complexity, or changing review norms; the PR-size data (Section 4.1.7) suggests that increasing PR scope is not the primary driver.

### 4.1.7 Metric 6: New-Author Share per Month

Figure 7 shows new vs. returning author counts per month, computed from the full merged-PR population. vLLM maintains an Era 3 average new-author share of approximately 40%; SGLang is at approximately 40%. Both projects sustain a mix of new and returning contributors across all eras, with the returning contributor base growing in absolute terms even as the new-author share gradually declines from earlier peaks (Era 0: ~61% for vLLM; Era 1: ~64% for SGLang). This pattern is consistent with a maturing community: the absolute number of new contributors continues to grow, but the returning contributor base grows faster, reflecting successful onboarding and retention [27, 28].

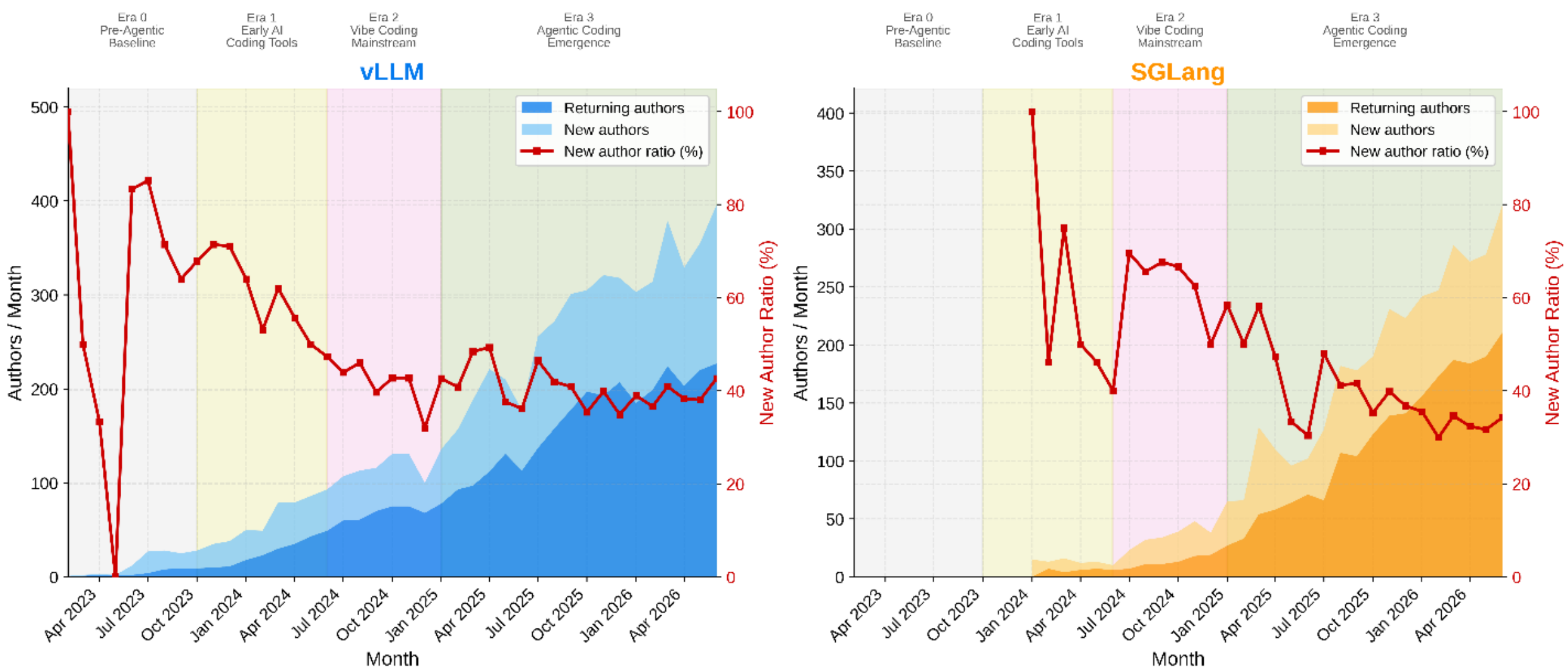

**Figure 7:** Stacked area chart of new vs. returning authors per month by project, with new-author share overlay, computed from the full merged-PR population. Both projects sustain ~40% new-author shares in Era 3, down from higher ratios in earlier eras, reflecting community maturation.

### 4.1.8 Metric 7: PR Size

Figure 8 shows median lines added, lines deleted, and files changed per merged PR over time, extracted from git commit statistics. This metric was added to test the hypothesis that rising comment density reflects increasing PR scope. The data show remarkable stability: vLLM's median additions per PR were 18 lines in Era 0 and 19 lines in Era 3; median files changed remained at 2 across all eras. SGLang shows a similar pattern: 26 median additions in Era 1 and 22 in Era 3, with median files changed at 2 throughout. These results indicate that the typical PR did not grow larger across AI tooling eras, despite the 21× and 17.9× increases in throughput. This finding has two implications. First, the throughput growth was achieved through more PRs of similar size, not larger PRs—consistent with a fine-grained, high-frequency contribution pattern. Second, rising comment density is not explained by increasing PR scope, suggesting that other factors (community growth, changing review norms, or AI-specific review dynamics) drive the increase in discussion per PR.

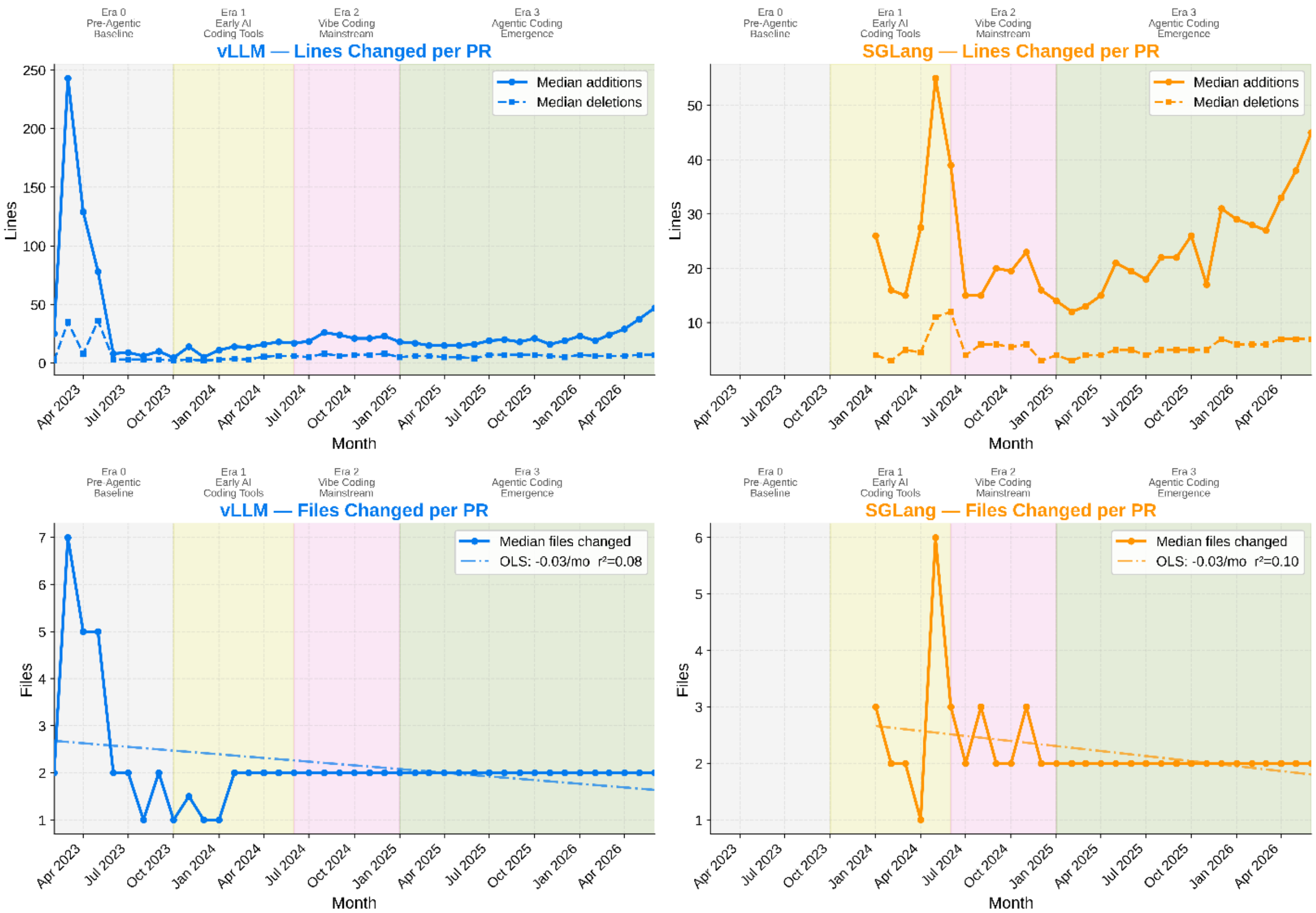

**Figure 8:** PR-size analysis: median lines changed (top panels) and median files changed (bottom panels) per merged PR over time, extracted from git commit statistics. PR size remained stable across eras despite order-of-magnitude throughput growth, indicating that throughput gains came from more PRs of similar scope rather than larger PRs.

### 4.1.9 Human vs. Bot Metric Comparison

Figure 9 presents a comprehensive comparison of human vs. bot activity across all metrics. The key observations are:

1. PR Throughput: Bot-authored PRs (dashed lines) are negligible compared to human-authored PRs (solid lines) in both projects. The bot contribution is so small that the dashed lines are barely visible above the x-axis.
2. Cycle Time: Bot PRs have highly variable cycle times (when they occur), but the sample size is too small to draw meaningful conclusions.
3. Unique Authors: Bot authors are a tiny fraction of unique authors (<1% in all eras).
4. Comment Density: Human-only comment density (solid lines) shows the same upward trend as total comment density, confirming that the increase is not driven by bot comments.
5. PRs per Author: Human PRs-per-author ratio remains stable across eras, indicating that individual productivity has not changed dramatically.
6. Bot PR Share: Bot PR share peaked in Era 2 at 0.32% (vLLM) and 0.07% (SGLang), then declined in Era 3 despite the overall throughput surge.

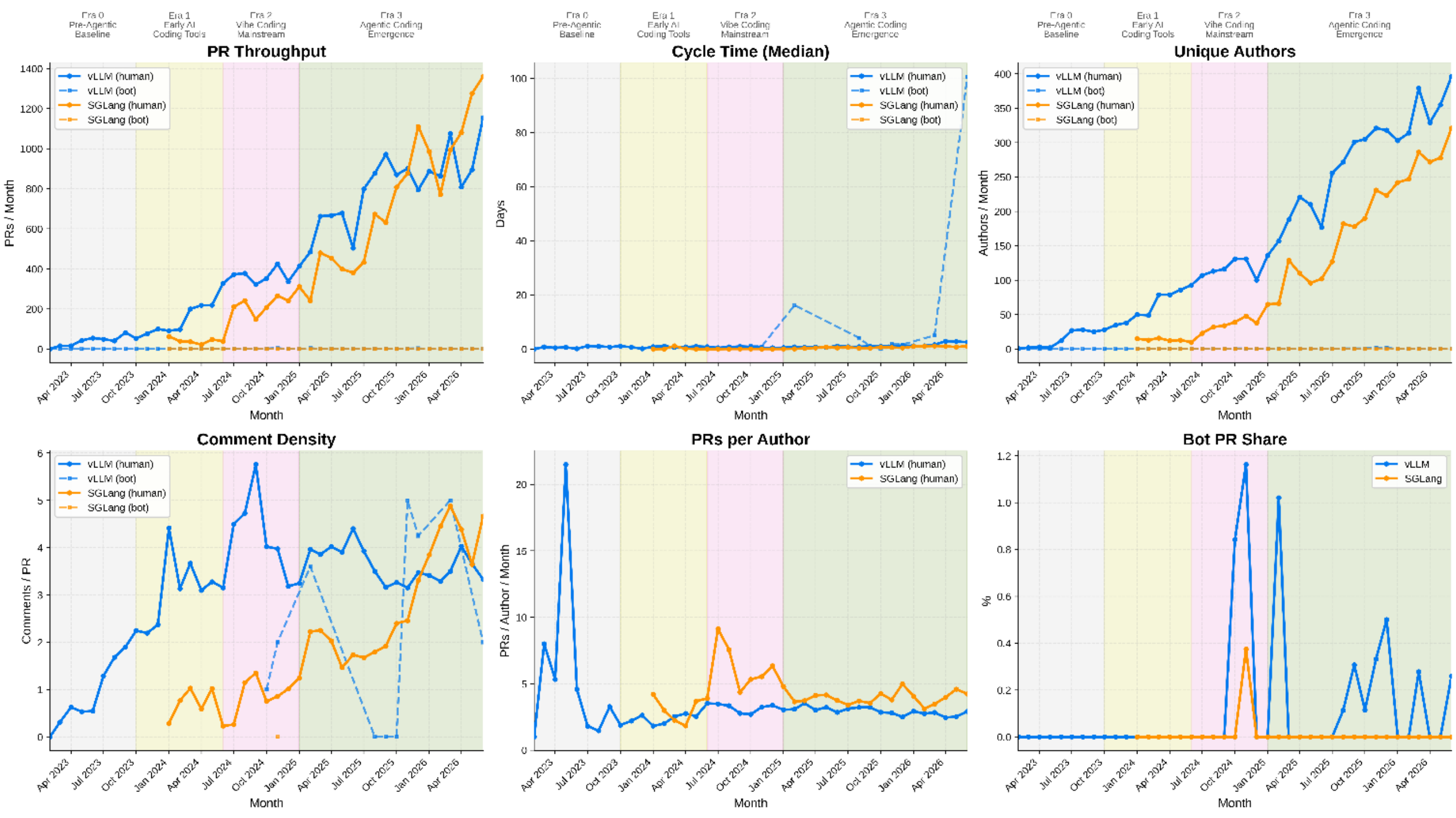

**Figure 9:** Human vs. bot metric comparison across six metrics. Solid lines show human activity; dashed lines show bot activity. Bot activity is negligible for PR throughput, unique authors, and PRs per author. Bot comments contribute an estimated 15–20% of the comment density increase. Bot PR share peaked in Era 2 and declined in Era 3.

# 5. Discussion

## 5.1 RQ #2: Detecting Shifts in Human-AI Collaboration

The seven metrics provide a quantitative foundation for describing changes in human-AI collaboration patterns. Several composite signals are particularly informative:

**Green flags** (potential indicators of healthy collaboration): Rising comment density concurrent with rising throughput may signal increased discussion and coordination around pull requests, although comment density alone cannot establish that teams are tackling more ambitious problems or reviewing them more rigorously. New contributors converting to returning contributors may indicate effective onboarding and successful integration into the project community. Stable PR size alongside rising throughput may suggest disciplined scoping and decomposition of changes.

**Warning signs** (indicators of collaboration drift): The Era 3 decline in monthly merge rate (vLLM: 0.685→0.541; SGLang: 0.703→0.620) suggests that the growing volume of opened PRs is outpacing merge capacity, which may indicate review bottleneck formation. The Era 3 increase in cycle time P90 (vLLM: 16.8 days; SGLang: 14.3 days) similarly suggests that complex PRs are waiting longer for review. A declining new author share with rising throughput may indicate that AI tooling is concentrating contribution rather than expanding it.

These signals are consistent with the CHAI-T framework for managing trust in human-AI collaboration [29], which emphasizes the importance of active trust calibration rather than passive acceptance of AI outputs. Dawarka et al. [30] similarly find that appropriately calibrated trust—mediated by transparency and role clarity—improves collaboration outcomes.

### 5.2 Leadership Framework: Seven Metrics, Seven Practices

Table 3 maps each metric to a concrete engineering management practice derived from the empirical findings.

**Table 3:** Leadership framework: seven metrics mapped to engineering management practices.

| Metric | Signal Type | Recommended Practice |
|---|---|---|
| **PR Throughput** | AI-assisted velocity | Set throughput targets by era; do not compare Era 3 teams to Era 0 baselines |
| **Cycle Time** | Review speed & PR scope | Monitor P90 separately from median; investigate tail latency as a review-capacity signal |
| **Unique Authors** | Community growth & onboarding | Invest in onboarding documentation; measure first-PR-to-second-PR conversion rate |
| **Comment Density** | Review engagement | Track rising density as a signal of community participation; investigate whether comments reflect clarification or architectural discussion |
| **Monthly Merge Rate** | Review capacity | Monitor for declining trends; a falling merge rate may indicate review bottleneck formation |
| **New-Author Share** | Community health & sustainability | Celebrate returning contributors; build mentorship paths for new authors |
| **PR Size** | Contribution scoping | Track median lines/files per PR; stable size with rising throughput indicates disciplined scoping |

### 5.3 The Judgment Gap

A central implication of the data is what this study terms the Judgment Gap: the widening space between what AI agents can execute and what engineering leadership requires. AI agents are effective at generating syntactically correct code, following established patterns, writing tests for specified behavior, and refactoring within defined scope [12, 13]. They are less effective at knowing when not to build a feature, detecting architectural drift, reading team morale signals, negotiating scope with stakeholders, or deciding which technical debt to carry. These are judgment calls that require organizational context, historical awareness, and sustained human relationships. As AI handles more execution, the premium on human judgment increases rather than decreases. Senior engineers who develop strong architectural judgment are more valuable in the agentic era, not less—but their value is no longer primarily in code output. This is consistent with findings that experienced developers exercise deliberate control over AI agent use [11] and that trust in AI-assisted code review is conditional and calibrated rather than unconditional [31].

### 5.4 Context Collapse

A second risk pattern is context collapse: AI agents generating code without the organizational context that makes it appropriate. The code may be technically correct, pass all tests, and follow style conventions, while simultaneously being architecturally wrong for the team, the codebase, or the current moment. Context collapse is not an AI failure—it is a leadership failure. Context is a leadership deliverable: architectural principles, team conventions, and current priorities must be written down, kept current, and made accessible to both AI agents and human contributors.

### 5.5 RQ #3: Mapping to Bioinformatics Engineering

The metrics and signals identified in this study may offer an illustrative mapping to bioinformatics pipeline development, with important domain-specific modifications. We emphasize that no bioinformatics teams were analyzed in this study; the mapping is conceptual and requires empirical validation. In bioinformatics, the pipeline is the experiment: a wrong parameter—incorrect genome build, inappropriate normalization method, or misspecified sample grouping—is not a software bug but a scientific error that may not surface for months, with consequences including retracted publications, misdiagnosed patients, or wasted grant funding.

The wet-lab-to-pipeline handoff introduces a multi-stage context collapse risk. A wet-lab scientist specifies biological intent; a bioinformatician translates that to computational logic; an AI agent generates pipeline code. Each translation step loses context. Recent AI bioinformatics systems—including PromptBio [33], ToolsGenie 2.0 [34], CARIBOU [35], the KBase Research Agent [36], and Biomni Lab [37]—demonstrate that end-to-end automation is technically feasible, but fully addressing the biological validity validation still requires Human-in-the Loop paradigm.

These AI bioinformatics agents rely on LLM inference engines for serving, creating an organizational analogy between the software-engineering processes studied here and the engineering challenges involved in building and maintaining high-performance inference systems. For example, Biomni Lab integrates LLM reasoning with retrieval-augmented planning and code-based execution, requiring low-latency and high-throughput inference to support interactive research workflows [37]. Our proof-of-concept benchmark comparisons show that SGLang achieves approximately 9% higher throughput than vLLM on smaller models and excels at structured output generation through its RadixAttention prefix caching, while vLLM demonstrates lower time-to-first-token latency and stronger single-request performance. The repository-level metrics examined in this study do not directly measure inference performance; rather, they provide a complementary perspective on the development processes and coordination dynamics underlying the engineering of such systems. The choice between these engines involves trade-offs in throughput, latency, and memory efficiency that directly impact the responsiveness and reliability of AI agents built on top of them.

Table 4 maps each engineering metric to its potential bioinformatics equivalent. Practical recommendations for bioinformatics development managers include: (1) redefining the handoff protocol to treat experimental design documents as first-class repository artifacts versioned alongside code; (2) building biological validation into CI/CD pipelines as the equivalent of automated unit tests; (3) maintaining a “bad vs. sad” classification for biological errors, distinguishing fixable computational errors from communication failures that require re-engagement with the wet-lab scientist; (4) treating data provenance as a first-class metric, given that AI-generated pipelines are particularly prone to provenance gaps; and (5) monitoring for the "plausible but wrong" failure mode, in which a pipeline runs without errors and produces results in the expected format, but the results are biologically invalid [32].

**Table 4:** Illustrative mapping of seven engineering metrics to bioinformatics pipeline development. This mapping is conceptual; no bioinformatics teams were analyzed in this study.

| Metric | SE Meaning | Bioinformatics Mapping |
|---|---|---|
| **PR Throughput** | AI-assisted velocity | Pipeline iteration speed: time to test a new normalization method or add a sample type |
| **Cycle Time** | Review speed | Validation turnaround: time from pipeline change to validated result on a test dataset |
| **Unique Authors** | Community growth | Cross-disciplinary collaboration: wet-lab scientists, clinicians, and bioinformaticians contributing to pipeline specs |
| **Comment Density** | Review engagement | Scientific scrutiny: biological assumptions challenged in code review, not just syntax |
| **Monthly Merge Rate** | Review capacity | Pipeline acceptance rate: fraction of proposed analyses surviving biological validation |
| **New-Author Share** | Community health | Knowledge transfer: wet-lab scientists becoming pipeline contributors; bioinformaticians learning the biology |
| **PR Size** | Contribution scoping | Pipeline change granularity: are changes focused and reviewable, or monolithic? |

### 5.6 Limitations

Several limitations of this study should be noted. First, the two repositories studied are not representative of all open-source projects: they are high-velocity, well-funded AI infrastructure projects with strong review cultures and active maintainers. The findings may not generalize to smaller projects, enterprise codebases, or projects with weaker review norms. Second, the era segmentation is based on the general availability dates of AI tooling milestones, which are proxies for actual adoption within these specific teams. Third, this study cannot establish causal attribution: the observed metric changes are correlated with AI tooling eras but may reflect confounding factors such as project maturity, funding changes, or community growth independent of AI tooling. Fourth, while we conducted a bot/human decomposition analysis using GitHub account-type metadata, our bot comment share estimates are based on a validation sample of 10 recent PRs and may not capture all bot activity patterns. Fifth, the comment density metric captures issue comments (general discussion thread) only and does not include inline review comments or formal review submissions; bot comments are included in the count, but our decomposition shows they contribute only 15–20% of the observed increase. Sixth, the monthly merge rate is a flow ratio comparing PRs merged in month t to PRs opened in month t; these are different cohorts, so the ratio does not represent an acceptance probability. Seventh, PR-size data were extracted from git commit statistics for squash-merged commits; 2.2% of merged PRs could not be matched to a commit and are excluded from PR-size analyses. Eighth, the GitHub Search API may not return all PRs for very high-volume months despite half-month splitting; we estimate coverage at >99% based on cross-validation with git commit counts.

## 6. Conclusion

This study presents a descriptive longitudinal case analysis of engineering dynamics across four eras of AI tooling adoption in two high-velocity open-source AI infrastructure projects, covering the complete population of merged pull requests (33,228 PRs). The central finding is that throughput, contributor participation, and comment density all increased substantially across the study period, while PR size remained stable. However, not all trends were favorable: monthly merge rate declined in Era 3 for both projects, and cycle time P90 increased, suggesting emerging review capacity constraints.

A bot/human decomposition analysis confirms that these trends are overwhelmingly human-driven. Botauthored PRs account for <0.2% of total merged PRs and contribute 0.17% (vLLM) and 0.00% (SGLang) of the observed throughput growth. Bot comments contribute an estimated 15–20% of the comment density increase, with the remaining 80–85% attributable to human activity. These findings rule out the hypothesis that automated bot activity explains the observed engineering metric changes.

The most notable finding—that PR comment density rose 4.2× in vLLM and 3.8× in SGLang across eras while PR size remained stable—raises the hypothesis that AI-era PRs attract more review discussion for reasons other than increased scope. This may reflect growing community participation, changing review norms, or AI-specific review dynamics. Distinguishing among these explanations requires comment-level analysis and is a priority for future work.

These findings highlight several implications for engineering organizations adopting AI-assisted development workflows. First, traditional productivity benchmarks may require recalibration as AI tooling changes the baseline scale of software contribution. Second, as code generation becomes increasingly automated, review, validation, and integration capacity may emerge as critical bottlenecks. The divergence between high throughput and persistent long-tail cycle times suggests that human coordination remains essential for complex engineering decisions. Third, maintaining explicit technical context—including architectural principles, development conventions, and domain-specific intent—may become increasingly important for effective collaboration among human developers and AI agents.

Future work should extend this analysis to a broader sample of repositories across different domains, team sizes, and AI tooling adoption levels. Causal identification strategies—such as difference-in-differences designs exploiting the staggered adoption of AI tools across teams—would strengthen the attribution of observed metric changes to AI tooling. Comment-level analysis (classifying comments as clarification, approval, or architectural discussion) would help explain the comment density increase. The bioinformatics mapping presented here warrants empirical validation through prospective studies of bioinformatics core teams adopting AI pipeline generation tools.

## Acknowledgments

The author thanks the maintainers of the vLLM and SGLang open-source projects for their transparent development practices and publicly accessible repository histories. Data were collected via the GitHub public Search API under standard rate limits. All analysis code is available upon request.

## Data Availability

All datasets are available upon request.

## Competing Interests

The author declares no competing interests.